\documentclass[twocolumn,showpacs]{revtex4}

\newcommand{\A}{{\bf A}}

\renewcommand{\k}{{\bf k}}

\newcommand{\p}{{\bf p}}
\renewcommand{\P}{{\bf P}}
\newcommand{\rr}{{\bf r}}

\newcommand{\Fig}[1]{Fig.~\ref{#1}}
\newcommand{\Figure}[1]{Figure~\ref{#1}}

\newcommand{\xy}{\textsl{XY} }

\usepackage{graphicx}

\begin{document}

\title{A Zero-Temperature Study of Vortex Mobility in\\
  Two-Dimensional Vortex Glass Models}

\author{Petter \surname{Holme}}
\author{Peter \surname{Olsson}}

\affiliation{Department of Theoretical Physics, Ume\aa\ University, 
  901 87 Ume{\aa}, Sweden}

\begin{abstract}
  Three different vortex glass models are studied by examining the energy
  barrier against vortex motion across the system. In the two-dimensional
  gauge glass this energy barrier is found to increase logarithmically with
  system size which is interpreted as evidence for a low-temperature phase
  with zero resistivity. Associated with the large energy barriers is a
  breaking of ergodicity which explains why the well established results
  from equilibrium studies could fail.  The behavior of the more realistic
  random pinning model is however different with decreasing energy barriers
  a no finite critical temperature.
\end{abstract}


\maketitle

The effect of disorder on the behavior of type-II superconductors in
magnetic fields is a subject of enormous interest both theoretically and
experimentally. Much discussion has the last few years focused on the
possibility of a stable vortex glass phase \cite{Fisher_Fisher_Huse} and
some recent papers find evidence from simulations for the existence of such
a phase in three dimensions
\cite{Olson_Young:00,Kawamura:00,Vestergren_Lidmar_Wallin}. The necessary
ingredients of a vortex glass model is, however, still an open question
\cite{Huse_Seung,Kawamura:00} and recent results actually suggest that two
of the popular models, the gauge glass \cite{Olson_Young:00} and the random
pinning model \cite{Vestergren_Lidmar_Wallin}, belong to different
universality classes.

In two dimensions, which is the focus of the present Letter, the different
vortex glass models are commonly believed to behave similarly.  One of these
is the gauge glass for which equilibrium
\cite{Nishimori:94,Fisher_Tokuyasu_Young,Gingras:92,Kosterlitz_Akino}
analyses show no transition.  In the present paper we approach this model
by examining the low-temperature properties of a vortex hopping dynamics.
The possibility we explore is that ergodicity breaking might lead to
different conclusions when considering dynamics as compared to equilibrium,
and we actually find results that strongly suggest a finite-temperature
transition.  Interestingly, the behavior of the random pinning model is
qualitatively different and this points to rich and unexpected behavior of
the vortex glass models even in two dimensions.

Results in statistical physics often rely on an assumption of ergodicity.
Only if this assumption is valid it is permissible to draw conclusions
regarding the physically relevant time-averages from the ensemble averages
\cite{Binder_Young}.  Generally speaking this assumption is expected to be
valid for models with a smooth energy landscape, whereas an energy
landscape with valleys separated by huge energy barriers, could give rise
to a breaking of ergodicity.  The concept ergodicity breaking presupposes
some kind of dynamics and in the present Letter we examine a vortex hopping
dynamics where one vortex moves one lattice constant at a time.  This is
the kind of dynamics that is typically used for the study of dynamical
properties in vortex glass models
\cite{Hyman_WFGY,Ying-Hong.Li:92,Kim:gauge-glass}.

The possibility of ergodicity breaking means that the standard arguments
for the absence of a phase transition could fail and this applies both to
the rigorous analytical argument by Nishimori \cite{Nishimori:94} and the
vanishing of the zero-temperature domain wall energy with increasing system
size \cite{Fisher_Tokuyasu_Young,Gingras:92,Kosterlitz_Akino}. The
dynamical studies should, however, still be reliable, but they arrive at
conflicting conclusions.  Whereas some have reported evidence for a zero
temperature transition \cite{Hyman_WFGY} others are strongly in favor of a
transition at a finite temperature \cite{Ying-Hong.Li:92,Kim:gauge-glass}.
The different conclusions seem to be due to different judgments regarding
the reliability of data at rather low temperatures.

In this Letter we present a novel zero-$T$ approach designed to measure
the height of the energy barriers in vortex glass models. The quantity
in focus is the energy barrier against phase slips (defined more
precisely below). Beside probing the possibility of ergodicity
breaking this quantity is physically relevant since the resistivity
in dynamical simulations is inversely proportional to the time between
phase slips. A phase slip energy barrier that diverges as
$L\rightarrow \infty$ would suggest both breaking of ergodicity and a
vanishing resistivity (= immobile vortices) which in turn implies a
superconducting low-temperature phase \cite{Kim_Minnhagen_Olsson}.

To determine the phase slip energy barrier we start in the ground state and
perform an exhaustive search of all relevant local moves, in the vortex
representation.  Considering the huge phase space of these models this
could seem like an impossible task and the only reason why it is at all
feasible is that it is sufficient to focus on some relatively few states
with rather low energy.  Our main result is that the phase slip energy,
$V_L$, in the gauge glass model grows logarithmically with system size, a
finding that we interpret as evidence for the existence of a
superconducting low-temperature phase.

The Hamiltonian for the random gauge \xy model is
\begin{equation}
  H = \sum_{\langle ij\rangle} U\left(\theta_i - \theta_j -
  A_{ij} - \frac{1}{L}\rr_{ij}\cdot {\bf\Delta}\right),
  \label{Hamiltonian}
\end{equation}
where $\theta_i$ is the phase at lattice point $i$ on a square lattice of
size $L\times L$ (where $L$ is in units of the lattice constant), the sum
is over nearest neighbors, and $A_{ij}$ is a randomly chosen vector
potential in the interval $[-r\pi,r\pi)$, where $r$ is the disorder
strength. The gauge glass model corresponds to the case $r=1$. In the above
Hamiltonian we also include the twist variable ${\bf\Delta} =
(\Delta_x,\Delta_y)$ \cite{FTBC} to allow for phase slips. The unit vector
$\rr_{ij}$ picks $\Delta_x$ or $\Delta_y$ depending on the direction of the
link $ij$.

The spin interaction is often taken to be a cosines function, but to
be able to take advantage of the direct relation to the vortex
representation we have instead chosen $U(\phi) = J\phi^2/2$. With the
frustration given by
\begin{displaymath}
  f_\rr = \frac{1}{2\pi}\nabla\times\A_\rr,
\end{displaymath}
and fluctuating twists \cite{FTBC} the vortex Hamiltonian becomes
\begin{equation}
  H^{\rm v} =  -\frac{1}{2}\sum_{\rr,\rr'} (q_\rr + f_\rr)
  G(\rr - \rr') (q_{\rr'} + f_{\rr'})
  \label{Hv}
\end{equation}
where the $q_\rr$ are unit charges, the factor $1/2$ compensates for the
double counting, and $G(\rr)$ is the lattice Green's function
\begin{displaymath}
  G(\rr) = \left(\frac{2\pi}{L}\right)^2 \sum_{\k\neq0}
  \frac{e^{i\k\cdot\rr} - 1}{2\cos k_x + 2\cos k_y - 4}.
\end{displaymath}

For the discussion of the phase slip energy barrier we start by
considering this quantity in the pure 2D XY model. The ground state is
trivially given by $q_\rr=0$ for all $\rr$. To consider vortex
transport we create a vortex pair with a negative vortex at the origin
and a positive at $(1,0)$, and then move the positive one to $(2,0)$,
$(3,0)$,\ldots $(L,0)$. The twist variable $\Delta_y$ will change
proportionally to make the current in the $y$-direction vanish.  Since
$(L,0)$ in a system with periodic boundary conditions is the same as
the origin we are then back again in the ground state but with the
twist variable $\Delta_y = -2\pi$.  (We use the term phase slip for
this kind of process even in the vortex representation.)  By these
steps we have transported unit vorticity across the system.  In this
simple case the energy for a configuration with the positive vortex at
$(x,0)$ is given by the lattice Green's function, $G(x,0)$, and the
phase slip energy barrier for the pure 2D XY model is the value of
this energy at the largest separation, $V_L = G(L/2,0)$.

For the following we note that each elementary move may be described by
a position $\rr$ and a unit dipole vector $\p$:
\begin{displaymath}
  \begin{array}{rcll}
  q_\rr - 1 & \rightarrow & q_\rr, & \mbox{put a $-1$ vortex at $\rr$},\\
  q_{\rr+\p} +1 & \rightarrow & q_{\rr+\p}, \quad & \mbox{put a $+1$ vortex
  at $\rr+\p$}.
  \end{array}
\end{displaymath}
To keep track of the transport of vorticity we introduce the
polarization $\P$. An elementary move always changes the polarization,
$\P + \p \rightarrow \P$.  The vorticity transport discussed above
starts in the ground state and ends in the same configuration but with
$\Delta P=(L,0)$. The polarization therefore contains some memory of
the steps taken and may be used to examine the phase slip.

\begin{figure}
  \includegraphics[width=8.2truecm]{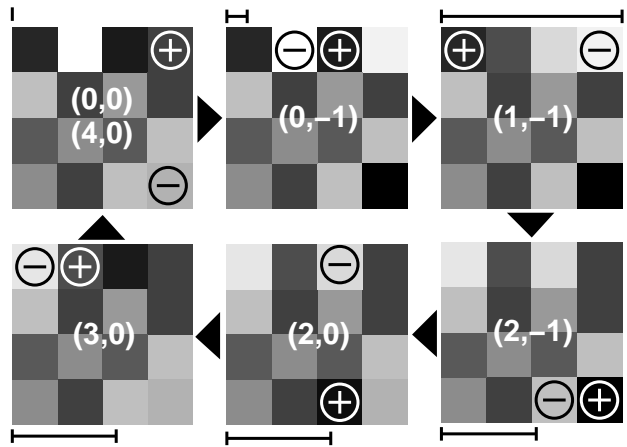}
  \caption{Illustration of a sequence of elementary moves for a
    realization of a $4\times 4$-system, corresponding to transport
    of vorticity across the system, $\Delta \P = (4,0)$.  The upper
    left corner is the ground state.  Charge is symbolized by the
    shading of the squares, black represents $-0.67$, white represents
    $+0.72$, and intermediate charges are proportionally grey. The digits
    show the polarization relative to the ground state. The charge
    symbols represent the dipole excited in the next step. The bar next
    to the square shows the energy of each configuration relative to
    the ground state.}
  \label{path}
\end{figure}

For the gauge glass model the sequence of elementary moves that corresponds
to the lowest possible energy barrier is in general much more complex than
in the pure XY model. To illustrate this point we show in \Fig{path} such a
sequence of moves for a certain disorder realization in a $4\times 4$
system.  In this case six steps are needed to transport vorticity across
the system and return back to the ground state.  We now describe the method
used to find such complex paths.

We first search for the ground state by applying the standard spin quench
algorithm \cite{Walker_Walstedt} a large number of times.  The lowest
energy state is taken to be the true ground state when it has been found
$N_{\rm rep} = 10$ times. No change was found in the results when we
instead used $N_{\rm rep} = 100$ for a set of $10^4$ disorder
configurations with $L=4$.  We then use an algorithm that is analogous to
the filling of a (energy) landscape with a liquid rising from a source at
the lowest position. At each time step it is the lowest level accessible by
the liquid that is invaded.  The disorder configurations give rise to
different periodic energy landscapes in polarization space, and the liquid
is made to rise until there is a connection between these periodic copies.
The landscape picture is, however, a great over-simplification.  For each
value of the polarization there is a large number of possible
configurations with different energy. The true picture is therefore more of
a plumber's nightmare with a large network of pipes criss-crossing the
polarization space but the intuitive picture still holds. With an
algorithm that slowly increases the level one is guaranteed that the first
connection found is the lowest possible.

In the computer, the algorithm consists of repeating the following
steps:
\begin{enumerate}
\item Generate $4L^2$ configurations by applying the $4L^2$ possible dipole
  excitations to the current configuration.
\item Calculate the energy of each such configuration and put them in a sorted
  list, lowest energy first, together with their polarization relative to the
  ground state.
\item Take the first (lowest energy) configuration from this list to be the new
  current configuration.
\item If this configuration has already been encountered, but with a
  different polarization such that $\Delta\P = (\pm L,0)$ then we are
  done. Otherwise, go to step~1.
\end{enumerate}

To make this algorithm work one also needs a list of already used
configurations. In step~1 the current configuration is added to that list
and in step~4 each configuration is compared to the list.  The main output
is the phase slip energy barrier which equals the highest energy used in
the algorithm. The data below are obtained by averaging the energy barrier
from 3000 to $2\times 10^5$ disorder configurations.  Much as expected, the
number of iterations, $N$, of the above algorithm grows rapidly with $L$,
$N \approx 0.5 \exp L$. In our simulations the spin quench algorithm that
is used to find the ground state is however more time consuming than the
determination of the energy barrier.

\begin{figure}
  \includegraphics[width=8.5truecm]{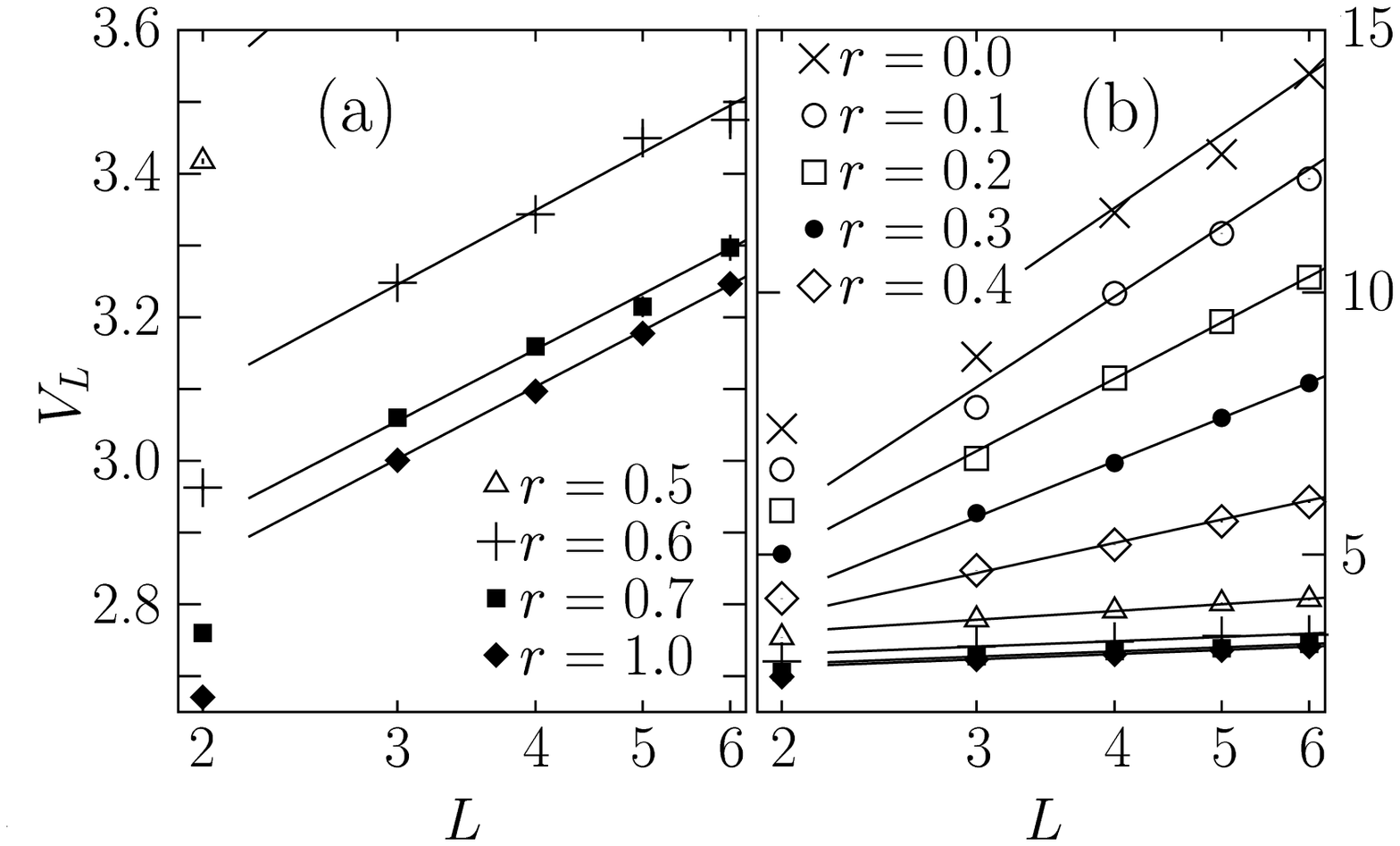}
  \includegraphics[width=8.5truecm]{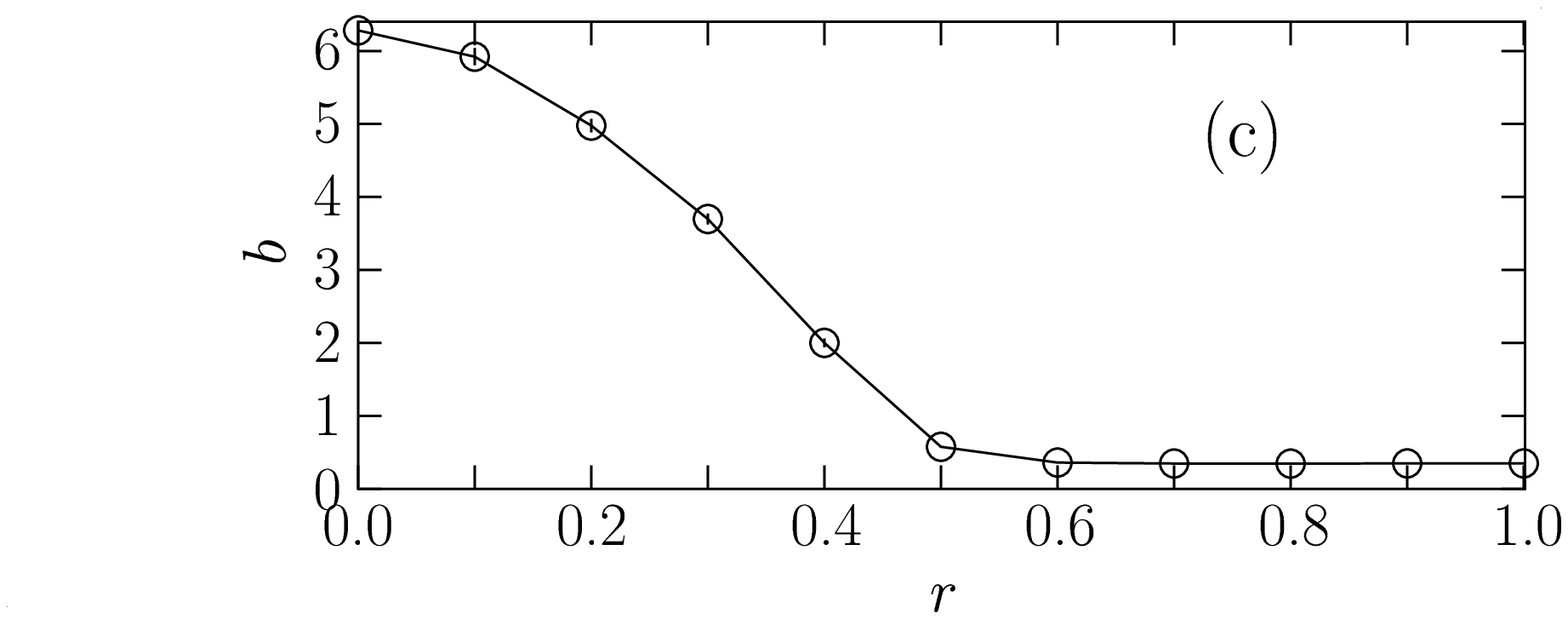}
  \caption{Size dependence of the phase slip energy $V_L$ for the
    random gauge \xy model. The result for the gauge glass ($r=1$) are
    shown by the solid diamonds in panel (a) whereas the results for the
    ordinary 2D \xy model ($r=0$) are shown by crosses in panel (b). We
    always find $V_L \sim \log L$ which suggests diverging energy barriers
    against vortex motion and a superconducting low-temperature phase.
    Panel (c) shows the slope of the lines in (b).  There are error bars
    corresponding to one standard deviation for all the data points, but
    most of them are far too small to be visible. }
  \label{Vl-gg}
\end{figure}

\Figure{Vl-gg} shows $V_L$ versus $L$ for several different values of
$r$. We first focus on the results for $r=1$, the usual gauge glass
model, shown by solid diamonds in panel (a).  By plotting the data
with log scale on the $x$ axis it is found that the points with
$L\geq3$ to an excellent approximation fall on a straight line with a
positive slope.  We consider this to be strong evidence for an energy
barrier that diverges with increasing lattice size and thereby also
evidence for the existence of a low-temperature phase with immobile
vortices.  As shown in the figure the scaling holds down to $L=3$.
This is the same size as in the examinations of the domain wall energy
in the gauge glass \cite{Kosterlitz_Akino} and it therefore seems likely
that the logarithmic increase in \Fig{Vl-gg} is the true behavior at
large $L$. As further support for this belief we note that the
logarithmic size-dependence is not a foreign behavior, but is rather
built into the model at the outset through the logarithmic
length-dependence in the vortex interaction.

For weaker disorder, $r<1$, the general feature is the same with $V_L \sim
\log L$.  Considering the change in behavior as $r$ decreases we find that
$V_L$ initially changes very slowly. The values of $V_L$ for $r=0.9$ and
0.8 (not shown) are almost identical to the ones for $r=1$. For an analysis
of the data for $r\leq 0.7$ we fit $V_L = a(r) + b(r)\ln L$, shown by solid
lines.  When $r$ is reduced below $\approx0.8$, $a(r)$ slowly increases but
the slope $b(r)$ at first remains unchanged, \Fig{Vl-gg}c.  The slope
actually to a good approximation remains constant down to $r=0.5$ where it
starts to increase and finally approaches $b(0)=2\pi$ of the pure 2D \xy
model.  We believe that the behavior of the slope $b$ shown in panel (c) is
related to the finding \cite{Kosterlitz_Simkin,Maucourt_Grempel:97} of a
phase with quasi long range order (without ergodicity breaking) for
$r<r_c\approx 0.37$. \Figure{Vl-gg}c is also very similar to the phase
diagram for the same model recently obtained on the basis of Monte Carlo
simulations \cite{Holme_Kim_Minnhagen}.

Since the gauge glass model is an often used model of a disordered
superconductor in an applied magnetic field, the above result would
quite surprisingly seem to suggest the existence of a dissipation-free
low temperature phase in disordered thin superconducting films. To
examine this question in more detail we have also studied the phase
slip energy in the random pinning model introduced in
Ref.~\cite{Hyman_WFGY}, which is meant to be a more realistic model of
a disordered superconductor in an applied magnetic field.  The
difference is that the random pinning model has a constant non-zero
magnetic field at each plaquette whereas the random gauge \xy model
instead is a superconductor with random fields that sum up to zero.
The Hamiltonian for the random pinning model is
\begin{equation}
  H_{\rm rp} = -\frac{1}{2}\sum_{\rr,\rr'} (q_\rr - f) G(\rr - \rr')
  (q_{\rr'} - f) -\sum_\rr v_\rr q_\rr^2.
  \label{Hrp}
\end{equation}
The frustration (magnetic field) is here homogeneous and the disorder
is instead included through the pinning potential $v_\rr$. We follow
Ref.~\cite{Hyman_WFGY} and let $v_\rr$ be a random variable
uniformly distributed between $-\pi$ and $\pi$, and restrict the
possible values for the vorticity to $q_\rr = -1, 0, 1$.

\begin{figure}
  \includegraphics[width=8.2truecm]{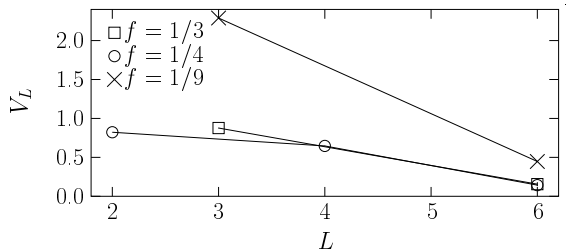}
  \caption{Size-dependence of $V_L$ for the random pinning model with
  three different values for the frustration. The energy barrier
  decreases with increasing $L$.}
  \label{Vl-rp}
\end{figure}

\Figure{Vl-rp} shows the size dependence of $V_L$ for the random pinning
model with $f=1/3$, $1/4$, and $1/9$.  Due to the requirement that $fL^2$
has to be an integer it is only possible to get data for a few lattice
sizes for each value of $f$. The behavior is here markedly different from
what we found in the gauge glass; $V_L$ decreases with increasing lattice
size and we therefore conclude that the phase slip energy barrier vanishes
in the thermodynamic limit. The results are therefore very different from
Ref.~\cite{Hyman_WFGY} where it was concluded from a scaling analysis of
the $I$-$V$ characteristics that the gauge glass model and the random
pinning model do belong to the same universality class. Interestingly,
recent simulations strongly suggest that the three-dimensional versions of
these two models belong to different universality classes
\cite{Olson_Young:00,Vestergren_Lidmar_Wallin}.

We also shortly mention our results for a generalized \xy spin glass,
which is given by the same Hamiltonian as the gauge glass but with
only two possible values for the vector potential, $A_{ij} = 0$,
$\pi$. Let $s$ denote the fraction of links with $A_{ij} =\pi$. Due to
the chiral symmetry every ground state has a chiral mirror image
$q_\rr \mapsto -q_\rr$ that is also a ground state.  We may therefore
define a chiral energy barrier $V_L^c$, as the energy barrier that has
to be climbed to reach the chirally mirrored ground state.

\begin{figure}
  \includegraphics[width=8.5truecm]{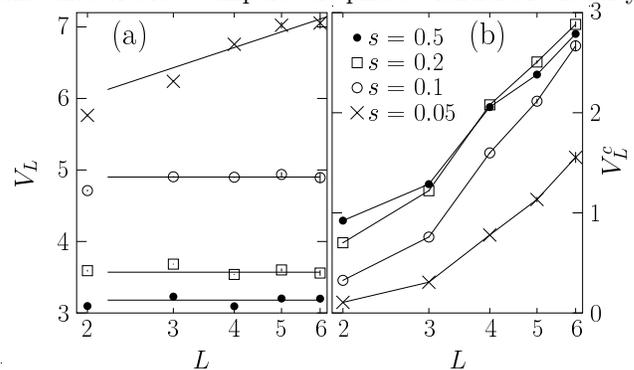}
  \caption{The energy barrier for $2\pi$ phase slip $V_L$ (a) and chiral
    mirroring $V_L^c$ (b) for the generalized $2\pi$ \xy spin glass. For the
    ordinary \xy model ($s=0$) $V_L$ is shown in \Fig{Vl-gg} while
    $V_L^c$ would be identically zero.}
  \label{xyspinglass}
\end{figure}

The phase slip energy $V_L$ is shown in \Fig{xyspinglass}a. For $s\geq
0$ $V_L$ is independent of size. For $s=0.05$ the data seems to go as
$\log L$ suggesting a critical $s_c>0.05$, but we cannot exclude the
possibility that $V_L$ saturates and approaches a constant for
$L\geq5$. The behavior of the chiral energy barrier in
\Fig{xyspinglass}b is different, suggesting $V_L^c \sim \log L$ for all
$s>0$ which means that the system at low temperatures and with only
local moves is trapped in the part of the configuration space with the
given chirality.

To conclude, we have examined the existence of growing energy barriers
and thereby a possible breaking of ergodicity in three different 2D
vortex glass models. By examining the phase slip energy barrier $V_L$
we found different behaviors in these three models. In the gauge
glass $V_L$ increases logarithmically which we take to suggest the
existence of a low-temperature phase with zero resistivity. In the
more realistic random pinning model $V_L$ instead decreases in
accordance with a transition at zero temperature. Yet another behavior
is found in the \xy spin glass where $V_L$ is independent of $L$.

We thank B. J. Kim, P.~Minnhagen, and S.~Teitel for valuable discussions.
This work was supported by the Swedish Natural Science Research Council
through Contract No.\ E 5106-1643/1999.


\end{document}